\immediate\write18{bibtex main}
\documentclass[runningheads,a4paper]{llncs}
\usepackage{amssymb,amsmath,makeidx,enumerate,stmaryrd,textcomp}
\usepackage[all]{xy}
\usepackage{tabls}
\usepackage{pslatex}
\usepackage{pdfsync}
\usepackage{hyperref}
\usepackage{color,graphicx,xcolor}
  \definecolor{brown}{rgb}{0.5,0,0}
  \definecolor{blue1}{rgb}{0.0, 0.0, 0.8}
  \definecolor{blue2}{rgb}{0.0, 0.0, 0.5}
\usepackage{hyperref}
  \hypersetup{
    colorlinks = true,
    urlcolor = {blue1},     
    linkcolor= blue2,       
    citecolor=blue2,        
    filecolor=blue2,         
    }

\usepackage[normalem]{ulem} 
\usepackage[T1]{fontenc}
\usepackage[utf8]{inputenc}
\DeclareUnicodeCharacter{2227}{$\land$}
\DeclareUnicodeCharacter{2228}{$\vee$}
\DeclareUnicodeCharacter{2208}{$\in$}
\DeclareUnicodeCharacter{22A0}{$\in$}
\DeclareUnicodeCharacter{22A2}{$\vdash$}
\DeclareUnicodeCharacter{2192}{$\rightarrow$}
\DeclareUnicodeCharacter{21D2}{$\Rightarrow$}
\DeclareUnicodeCharacter{27F9}{$\Longrightarrow$}
\DeclareUnicodeCharacter{03BB}{$\lambda$}
\DeclareUnicodeCharacter{27F8}{$\Longleftarrow$}
\DeclareUnicodeCharacter{209B}{$\textsubscript{S}$}
\DeclareUnicodeCharacter{209C}{$\textsubscript{F}$}
\DeclareUnicodeCharacter{2203}{$\exists$}

\usepackage{verbments}

\makeatletter
\newcommand{\verbments@write@detok}[1]{%
  \immediate\write\FV@OutFile{\detokenize{#1}}}
\newcommand{\verbments@FVB@VerbatimOut}[1]{%
  \@bsphack
  \begingroup
  \FV@UseKeyValues
  \FV@DefineWhiteSpace
  \def\FV@Space{\space}%
  \FV@DefineTabOut
  \let\FV@ProcessLine\verbments@write@detok
  \immediate\openout\FV@OutFile #1\relax
  \let\FV@FontScanPrep\relax
  \let\@noligs\relax
  \FV@Scan}
\let\FVB@VerbatimOut\verbments@FVB@VerbatimOut
\makeatother

\usepackage{bussproofs}

\newcommand{\RESagaProxy}{\,\rotatebox[origin=c]{90}{$\{\rotatebox[origin=c]{-90}{$\aga$}\}$}\,}

\newcommand{\mc}{\multicolumn}

\newcommand{\RESalphaProxy}{\,\rotatebox[origin=c]{90}{$\{\rotatebox[origin=c]{-90}{$\alpha$}\}$}\,}

\newcommand{\RESalphaBox}{\,\rotatebox[origin=c]{90}{$[\mkern1.8mu\rotatebox[origin=c]{-90}{$\alpha$}\mkern1.8mu]$}\,}

\newcommand{\RESalphaDia}{\,\rotatebox[origin=c]{90}{$\langle\rotatebox[origin=c]{-90}{$\alpha$}\rangle$}\,}

\def\aga{\texttt{a}}

\newcommand{\RESagaDia}{\,\rotatebox[origin=c]{90}{$\langle\rotatebox[origin=c]{-90}{$\aga$}\rangle$}\,}

\newcommand{\RESagaBox}{\,\rotatebox[origin=c]{90}{$[\mkern1.8mu{\rotatebox[origin=c]{-90}{$\aga$}}\mkern1.8mu]$}\,}

\newcommand{\ls}{\lbrack}
\newcommand{\rs}{\rbrack}
\newcommand{\lc}{\langle}
\newcommand{\rc}{\rangle}

\newcommand{\pand}{\wedge}

\def\aol{\rule[0.5865ex]{1.38ex}{0.1ex}}

\def\pdra{\mbox{$\,{\rotatebox[origin=c]{-90}{\raisebox{0.12ex}{$\pand$}}{\mkern-2mu\aol}}\,$}}
\def\pdla{\mbox{\rotatebox[origin=c]{180}{$\,{\rotatebox[origin=c]{-90}{\raisebox{0.12ex}{$\pand$}}{\mkern-2mu\aol}}\,$}}}

\def\fCenter{{\mbox{$\ \vdash\ $}}}

\EnableBpAbbreviations

\newcommand{\commment}[1]{}

\DeclareGraphicsExtensions{.pdf,.png,.jpg,.eps}

\begin{document}

\mainmatter

\title{Tool support for reasoning in display calculi}

\titlerunning{Reasoning in display calculi}

\author{Samuel Balco \inst{1}\thanks{Supported by an EPSRC-funded Vacation Bursary in Summer 2014 and the University of Leicester Career Development Service Graduate Gateways programme in Summer 2015.} 
\and 
Sabine Frittella \inst{2} 
\and 
Giuseppe Greco \inst{2}
\and 
Alexander Kurz \inst{3}
\and 
Alessandra Palmigiano\inst{2}\inst{4}}

\authorrunning{Balco \and Frittella \and  Greco \and  Kurz \and  Palmigiano}

\institute{
Department of Computer Science, University of Oxford 
\and
Faculty of Technology, Policy and Management, Delft University of Technology
\and
Department of Computer Science, University of Leicester 
\and
Department of Pure and Applied Mathematics, University of Johannesburg
}

\maketitle

\begin{abstract} 
We present a tool for reasoning in and about propositional sequent calculi. One aim is to support reasoning in calculi that contain a hundred rules or more, so that even relatively small pen and paper derivations become tedious and error prone. As an example, we implement the display calculus D.EAK of  dynamic epistemic logic. Second, we provide embeddings of  the calculus in the theorem prover Isabelle for formalising proofs about D.EAK. As a case study we show that the solution of the muddy children puzzle is derivable for any number of muddy children. Third, there is a set of meta-tools, that allows us to adapt the tool for a wide variety of user defined calculi.
\end{abstract}


\section{Introduction}

\medskip\noindent\textbf{Applied logic. }
An important development in logic, and in particular of logic in computer science, has been the move away from logic to logics. The need for automation of reasoning in AI or verification led to the design of hundreds of bespoke logics with good algorithmic properties for particular tasks. This development is particularly conspicuous in modal logic, the classic instance of which, Kripke's modal logic \textbf{K}, is decidable just because it is a certain fragment of first-order logic \cite{mlhandbook:BB}. To compensate for the lack of expressiveness that comes with decidability, one develops modal logics focussed on different aspects such as time, probability, space, etc. See \cite{mlhandbook} for an overview of examples and techniques.


\medskip\noindent\textbf{The proof theory of modal logic} has had many successes, for example the tableaux methods of description logics with its applications to knowledge representation and ontologies \cite{mlhandbook:F,mlhandbook:HHSS,mlhandbook:BL}.
Nevertheless, for many of the more sophisticated modal logics good proof systems are not known. A notable exception is the cut free sequent calculus of \cite{Dyckhoff} for  dynamic epistemic logic  (without common knowledge). But given the diversity of modal logics and the importance of proof systems for reasoning about applications, it is desirable to have a systematic and uniform approach to the construction of modal proof systems with good proof theoretic properties.  

\medskip\noindent\textbf{Display calculi. } Following work on so-called display calculi \cite{Belnap,Kracht,Gore96,Wansing,Gore98}, we engaged
in the systematic study of display calculi of dynamic modal logics in \cite{DEAK,Multitype,PDL}. The principal advantage of display calculi is that they are built in a modular way and important proof theoretic properties such as cut-elimination are preserved under combination of logics (if the combined logic is still displayable). 

\medskip\noindent\textbf{Dynamic Epistemic Logic. } As a case study for the feasibility of this approach we developed a display calculus D.EAK for the  dynamic epistemic logic \cite{Ditmarsch} of Baltag-Moss-Solecki \cite{bms} without common knowledge. On the one hand this logic contains features that are a challenge from a proof theoretic point of view. On the other hand, dynamic epistemic logic has many applications, both in computer science and in other areas. One particular interest is the verification of security protocols that involves epistemic notions. In this paper, as a case study of intermediate complexity, we give a full proof of the muddy children puzzle. A nice collection of more epistemic puzzles is available in \cite{DitmarschK15}.

\medskip\noindent\textbf{Modularity. }
One aspect of the modularity of display calculi is that the logic is axiomatised by the structural rules, which can be added or removed in a flexible way. For example, even though D.EAK is based on classical propositional logic it can as well be based on intuitionistic or substructural logics by removing some of the structural rules. Another aspect of modularity is that it is easy to combine different such display calculi. But modularity also comes at cost. For example, D.EAK has more than a hundred rules. This poses no conceptual problems as the space of rules is well structured according to clear proof theoretic principles, but it does pose the practical problem of conducting the proofs and writing them down without making mistakes. Consequently, already for proof theoretic studies alone, tool support will be valuable.

\medskip\noindent\textbf{Contributions. }
One aim of the tool is to support researchers working on the proof theory of display calculi. The typical derivations may by relatively small, but they must be presented in a user interface in Latex in a style familiar to the working proof theorist. Moreover, in order to facilitate experimenting with different rules and calculi, meta-tools are needed that construct a calculus toolbox from a  given calculus description file.

The second aim is to support investigations into the question whether a calculus is suited to reasoning in some application area. To perform relevant case studies, one must deal with much bigger derivations and additional features such as abbreviations and derived rules are necessary. Another challenge is that applications may require additional reasoning outside the given calculus, for which we provide an interface with the theorem prover Isabelle.  

More specifically, in the work presented in this paper, we focus on D.EAK and aim for applications to epistemic protocols. In detail, we provide the following.

\begin{itemize}
\item A calculus description language that allows the specification of the  terms and rules as well as of their typesetting in ASCII, Isabelle and LaTeX in a calculus description file.
\item A program creating from a calculus description file the calculus toolbox, which comprises the following.
\begin{itemize}
\item A shallow embedding of the calculus in the theorem prover Isabelle. The shallow embedding encodes the terms and the rules of the calculus and allows us to verify in the theorem prover whether a sequent is derivable in D.EAK.
\item A deep embedding of the calculus in Isabelle. The deep embedding also has a datatype for derivations and allows us to prove theorems about derivations. 
\item A user interface (UI) that supports
\begin{itemize}
\item interactive creation of proof trees,
\item simple automatic proof search (currently only up to depth 5),
\item export of proof trees to LaTeX and Isabelle,
\item the use of derived rules, abbreviations, and tactics.
\end{itemize}
\end{itemize}
\item A full formalisation of the proof system for dynamic epistemic logic of \cite{DEAK}, which is the first display calculus of the logic of Baltag-Moss-Solecki \cite{bms} (without common knowledge).
\item A fully formal proof (implemented in Isabelle) of the muddy children puzzle.
\item A set of meta-tools that enables a user to change the calculus.
\end{itemize}

\medskip\noindent\textbf{Case study: Muddy children}
 The first version of the tool presented at ALCOP 2015 supported the interactive construction of the proof trees in \cite{DEAK} and their output to \LaTeX. These proofs are not longer than a few dozen steps and of interest to establish  the mathematical result of completeness of the D.EAK. 
 
The muddy children puzzle was chosen because it is a well-known example of an epistemic protocol and required us to extend the tool from one supporting short proofs of theoretical value to larger proofs in an application domain.
 
On the UI side, we added features including abbreviations, macros (derived rules), and two useful tactics. On the Isabelle side, we added a shallow embedding of D.EAK in which we do the inductive proof that the well-known solution of the muddy children puzzle holds for arbitrary number of children. Whereas most of the proof is done in D.EAK using the UI and then automatically translating to Isabelle, the induction itself is based on the higher order logic of Isabelle/HOL.

\medskip\noindent\textbf{Related work. } The papers \cite{DawsonG01,DawsonG02} on proving cut elimination of display calculi in Isabelle have been a source of inspiration. Indeed, proving in Isabelle the variations \cite{DEAK,Multitype,PDL} of Belnap's cut-elimination theorem remains one of our aims and we consider what we present in this paper as a necessary first step: Due to the notational overhead resulting from encoding the mathematical description of D.EAK into its  Isabelle formalisation, we found constructing even the simplest derivations in Isabelle too burdensome without tool support.

\medskip\noindent The papers \cite{Lescanne06,LescanneP07,Lescanne13} implement epistemic logic in the proof assistant Coq. It would be very interesting to conduct the work of this paper based on Coq to enable an in depth comparison.

\medskip\noindent\textbf{Comparison of Isabelle to other proof assistants. } Isabelle has the following advantages for us.
\begin{enumerate}
\item Isabelle supports the proof language Isar supporting a style of writing proofs that is close to mathematical practice. 
\item Isabelle provides the so-called sledgehammer method, which uses specialised automatic theorem provers that are able to discharge much of the tedious, low level reasoning. 
\item Isabelle can export theories into programming languages such as Scala. This allows us to build the user interface directly on the deep embedding of the calculus in Isabelle, thus reusing verified code.
\end{enumerate}
As far as we know Isabelle is the only proof assistant featuring all of the above. This will be important to us in Section~\ref{sec:mc}, where we use (1) and (2) in order to write the mathematical parts of the proof of the solution of the muddy children puzzle in a mathematical style close to \cite{MaPS14} and we use (3) and the user interface to build the derivations in D.EAK.

\medskip\noindent\textbf{Outline. } Section~\ref{sec:deak}  reviews what one needs to know about D.EAK. Section~\ref{sec:deak-tool} presents the main components of the DEAK calculus toolbox. Section~\ref{sec:mc} discusses the implementation of the muddy children puzzle. Section~\ref{sec:meta-tool} explains the efforts we have made to keep the tool parametric in the calculus. Section~\ref{sec:conclusion} discusses directions of future research we plan to pursue.

\medskip\noindent\textbf{Acknowledgements. }  At several crucial points, we profited from expert advice on Isabelle by Tom Ridge, Thomas Tuerk and Christian Urban. We thank Roy Crole and Hans van Ditmarsch for valuable comments on an earlier draft.

\section{The display calculus D.EAK}\label{sec:deak}
This section gives some background on D.EAK; for a complete description we refer to \cite{DEAK} (where it is called D'.EAK). 

\newcommand{\Fma}{\phi}
\medskip\noindent\textbf{Formulas.} D.EAK is a proof system for (intuitionistic or classical) dynamic epistemic logic the formulas of which are defined by induction as follows:
\begin{equation}\label{eq:fma}
\begin{tabular}{|c|}
\hline
$\ \Fma ::= \textsf{AtProp} \mid \bot \mid \top \mid \Fma \wedge \Fma \mid \Fma \vee \Fma \mid \Fma\rightarrow\Fma \mid [\aga]\Fma \mid [\alpha]\Fma \mid \langle \aga\rangle\phi \mid \langle \alpha\rangle\phi \mid 1_\alpha\ \ $\\
\hline
\end{tabular}
\end{equation}
where $\aga$ ranges over agents, with $[\aga]\phi$ standing for ``agent $\aga$ knows $\phi$'', and $\alpha$ ranges over actions with $[\alpha]\phi$ standing for ``$\phi$ holds after $\alpha$''. $1_\alpha$ represents the precondition of the action $\alpha$ in the sense of \cite{bms}. Negation is expressed by $\Fma\rightarrow\bot$.

\medskip\noindent\textbf{Operational rules. } Display calculi are sequent calculi in which the rules follow a particular format that guarantees good proof theoretic properties such as cut elimination. One of the major benefits is modularity: different calculi can be combined and rules can be added while the good properties are preserved.

The rules of the calculus are formulated in such a way that, in order to apply a rule to a formula, the formula needs to be `in display'. For example, the following or-introduction on the left (where contexts are denoted by $W,X,Y,Z$ and formulas by $A,B$)
\begin{equation}\label{eq:or-seq}
\text{
{\small{
\AX$W,A \fCenter X$
\AX$Z,B \fCenter Y$
\LeftLabel{$\ (\vee_L')$ \ }
\BI$W,Z,A \vee B \fCenter X \,, Y$
\DisplayProof \\
}}}
\end{equation}
is not permitted in a display calculus, since the formula $A\vee B$ must be introduced in isolation as, for example, in our rule
\begin{equation}\label{or-deak}
\text{
{\small{
\begin{tabular}{|c|}
\hline
\AX$A \fCenter X$
\AX$B \fCenter Y$
\LeftLabel{$\ \ (\vee_L)\ $}\RightLabel{\ \ }
\BI$A \vee B \fCenter X \,; Y\ $
\DisplayProof \\
\hline
\end{tabular}
}}}
\end{equation}

\medskip\noindent\textbf{Display rules. } In order to derived a rule such as $(\vee_L')$ from the rule $(\vee_L)$, it becomes necessary to isolate formulas by moving contexts to the other side. This is achieved, by pairing the structural connectives such as ``,'' (written `;' in D.EAK) with so-called adjoint (aka residuated) operators such as ``$>$'' and adding bidirectional display rules
\begin{center}
{\small{
\begin{tabular}{|rl|}
\hline
\AX$X\, ; Y \fCenter Z$
\LeftLabel{\ $(;, >)$}
\doubleLine
\UI$Y \fCenter X > Z$
\DisplayProof & \ \ \ \ \ \ 
\AX$Z \fCenter X\, ; Y$
\RightLabel{$(>, ;)$\ }
\doubleLine
\UI$X > Z \fCenter Y$
\DisplayProof \\
\hline
\end{tabular}}}
\end{center}
which, in this instance, allow us to isolate $Y$ on the left or right of the turnstile. 

\medskip\noindent 
The name display calculus derives from the requirement that in a display calculus the so-called display property needs to hold: Each substructure can be isolated on the left-hand side, or, exclusively, on the right-hand side. This is the reason why we can confine ourselves, without loss of generality, to the special form of operational rules discussed above.

\medskip\noindent\textbf{Structures. }
A systematic way of setting this up for the set of formulas \eqref{eq:fma} is to introduce structural connectives corresponding to the operational connectives as follows. 
\begin{center}
\renewcommand{\arraystretch}{1.5}
\begin{tabular}{|c|c|c|c|c|c|c|c|c|c|c|c|c|c|c|c|c|c|c|}
 \hline
 \scriptsize{Structural} & \mc{2}{c|}{$<$}   & \mc{2}{c|}{$>$} & \mc{2}{c|}{$;$} & \mc{2}{c|}{I}  
 &\mc{2}{c|}{$\{\alpha\}$}       & \mc{2}{c|}{$\RESalphaProxy$} & \mc{2}{c|}{$\Phi_\alpha$}   
 &\mc{2}{c|}{$\{\aga\}$}       & \mc{2}{c|}{$\RESagaProxy$}   \\
 \hline
 \scriptsize{\ Operational\ } & $\, \pdla\, $ & $\, \leftarrow\, $ & $\, \pdra\, $ & $\, \rightarrow\, $  & $\, \wedge\, $ & $\, \vee\, $  & $\, \top\, $ & $\, \bot\, $    
 &  $\, \lc\alpha\rc\, $ & $\, \ls\mkern2mu\alpha\mkern1mu\rs\, $  & $\, \RESalphaDia\, $ & $\, \RESalphaBox\, $  & $\, 1_\alpha\, $ & $\, \phantom{1_\alpha}\, $
 &  $\, \lc\aga\rc\, $ & $\, \ls\mkern1.8mu{\aga}\mkern1.8mu\rs\, $   & $\, \RESagaDia\, $  & $\, \RESagaBox\,  $\\
\hline
\end{tabular}
\end{center}
This leads to a two tiered calculus which has formulas and structures, with structures generalising contexts and being built from structural connectives. 

We briefly comment on the particular choice of structural connectives above.
Keeping with the aim of modularity, D.EAK was designed in such a way that one can drop the exchange rule for `;' and treat non-commutative conjunction and disjunction. This means that we need two adjoints of `;' denoted by $>$ and $<$.
\footnote{
For example, taking into account the correspondence between operational and structural connectives, the rule $(;,>)$ above says precisely that the operation that maps $C$ to $A\rightarrow C$ is right-adjoint to the operation that maps $B$ to $A\wedge B$. Similarly, $(>,:)$ expresses that  $A\pdra \_$ is left adjoint to $A\vee\_$.
} 
Following the symmetries inherent in this substructural analysis of logic suggests to add the operational connectives $\pdla$, $\leftarrow$, $\pdra$, but they are not needed in the following. Similarly, the modal operators $[\alpha]$ and $[\aga]$ have structural counterparts $\{\alpha\}$ and $\{\aga\}$ which in turn have adjoints $\RESalphaProxy$ and $\RESagaProxy$. 
The formulas \eqref{eq:fma} do not have operational connectives corresponding to the structural connectives $\RESalphaProxy$ and $\RESagaProxy$, but they can be added and are indeed useful (in terms of Kripke semantics, the adjoint of a box modality $\Box$ for a relation $R$ is the diamond modality for the converse relation $R^{-1}$ often denoted by $\blacklozenge$).

\medskip\noindent\textbf{Structural rules. } 
The rules of D.EAK can be divided into operational rules and display rules, as discussed above, and structural rules, to which we turn now. The operational rules such as $(\vee_L)$ specify how to introduce a logical operation. Display rules such as $(;>)$  are used to isolate formulas or structures to which we want to apply a specific rule. The logical axiomatisation sits in the structural rules. Apart from the structural rules like weakening, exchange, and contraction for `;' we have also other structural rules such as the display rules discussed above and rules that express properties such as `actions are partial functions' axiomatised by the rule

\begin{equation}%
\text{
{\small{
\begin{tabular}{|c|}
\hline
\AX$X \fCenter Y$
\RightLabel{\emph{}}
\UI$\{\alpha\} X\fCenter \{\alpha\} Y$
\DisplayProof\\
\hline
\end{tabular}
}}}
\end{equation}
and such as `if $\aga$ knows $Y$, then $Y$ is true' axiomatised by
\begin{equation}%
\text{
{\small{
\begin{tabular}{|c|}
\hline
\AX$X \fCenter  \{\aga\} Y$
\RightLabel{\emph{}}
\UI$X\fCenter Y$
\DisplayProof\\
\hline
\end{tabular}
}}}
\end{equation}

\medskip\noindent\textbf{Modularity of D.EAK and related calculi. }  We have seen that D.EAK  has a large number of connectives. But they arise according to clear principles: operational connectives have structural counterparts which in turn have adjoints. Similarly, the fact that D.EAK has over a hundred rules poses no problems from a conceptual point as the rules fall into clearly delineated classes each serving their own purpose. It is exactly this feature which enables the modularity of the display logic approach to the proof theory of sequent calculi. But, from a practical point of view of creating proof trees or of composing a number of different calculi, this large number of connectives and rules makes working with these calculi difficult. Moreover, the  encoding of terms and proof trees needed for automatic processing will not be readable to humans who would expect to manipulate latexed prooftrees in an easy interactive way. How we propose to solve these problems will be discussed in the next section.

\section{The DEAK calculus toolbox}\label{sec:deak-tool}

The aim of the DEAK calculus tool \cite{Balco:DEAK} is to support research on the proof theory of dynamic epistemic logic as well as to conduct case studies exploring possible applications.  It provides a shallow and a deep embedding of D.EAK into Isabelle and a user interface implemented in Scala. 

The shallow embedding has an inductive datatype for the terms of the calculus and encodes the rules via a predicate describing which terms are derivable. It is used to prove correct the solution of the muddy children puzzle in Section~\ref{sec:mc}.

The deep embedding also has datatypes for rules and derivations and provides functionality such as rule application (match and replace) as well as automatic proof search and tactics. The corresponding Isabelle code is exported to Scala and used in the user interface.

D.EAK proof trees can be constructed interactively in a graphical user interface  by manipulating trees typeset in LaTeX. Proof trees can be exported to LaTeX/pdf and Isabelle.   This was essential for creating the Isabelle proof in Section~\ref{sec:mc}. Examples can be found in the the folder \texttt{LaTeX} in \cite{Balco:MC}: The \texttt{.cs} files contain the proofs as done in the UI and the \texttt{.tex}-files the exported LaTeX code. The tag \texttt{cleaned\_up} was added after a small amount of manual post-processing of the \texttt{.tex}-files.

\subsection{Shallow embedding (SE)  in Isabelle}\label{sec:se}

The shallow embedding of the calculus D.EAK is available in the files \texttt{DEAK\_SE.thy} and \texttt{DEAK\_SE\_core.thy}. The file \texttt{DEAK\_SE\_core.thy} contains the definitions of the terms via datatypes \texttt{Atprop, Formula, Structure, Sequent}. For example,

\begin{pyglist}[language = isabelle]
datatype Sequent = Sequent Structure Structure ("_ ⊢ _")
\end{pyglist}

\noindent declares that an element of datatype \texttt{Sequent} consists of two structures. The annotation \texttt{("\_ ⊢ \_")} allows us to use the familiar infix notation $\vdash$ in the Isabelle IDE. 

The file \texttt{DEAK\_SE.thy} encodes the rules of the calculus by defining a predicate \texttt{derivable}

\begin{pyglist}[language = isabelle]
inductive derivable :: "Locale list => Sequent => bool"  ("_ ⊢d _")
\end{pyglist}
\noindent by induction over the rules of D.EAK. For example, the rule $(\vee_L)$ above is encoded as

\begin{pyglist}[language = isabelle]
Or_L:  "l ⊢d (B ⊢ Y) ⟹ l ⊢d (A ⊢ X) ⟹ l ⊢d (A ∨ B ⊢ X ; Y)"
\end{pyglist}
\noindent which expresses in the higher-order logic of Isabelle/HOL that if $B\vdash Y$ and $A\vdash X$ are derivable, then $A\vee B\vdash X;Y$ is derivable.
Note that $A,B,X,Y$ are variables of Isabelle. The rule will be applied using the built-in reasoning mechanism of Isabelle/HOL which includes pattern matching. 

The datatype \texttt{Locale} is used to carry around all the information needed in a proof that is not directly available, in a bottom up proof search, from the sequent on which we want to perform a rule.

For example, in order to perform a cut, we need to specify the cut formula. In the UI, when constructing a prooftree interactively, it will be given by the user. Internally, cut-formulas are of type \texttt{Locale}  and the cut-rule is given by

\begin{pyglist}[language = json]
"(CutFormula f)∈set l ⟹ l⊢d(X ⊢ f) ⟹ l⊢d(f ⊢Y) ⟹ l⊢d(X ⊢Y)"
\end{pyglist}

Similarly, the rules that describe the interaction of the knowledge of agents with epistemic actions depend on the so-called action structures, which define the actions, but are not part of the calculus itself. These action structures, therefore, are also encoded by data of type \texttt{Locale}.

Before coming to the deep embedding next, we would like to emphasise, that in order to prove in the shallow embedding, that a certain sequent is derivable in D.EAK, one shows in theorem prover Isabelle/HOL that the sequent is in the extension of the predicate \texttt{derivable}. The proof itself is not available as data that can be manipulated. For example, with the shallow embedding, it will not be possible to write an Isabelle function that transforms a proof into a cut-free proof. (Cut-elimination is a topic we had to defer to future work, but we do make use of the deep embedding in the user interface.) 

\subsection{Deep embedding (DE) in Isabelle}

The deep embedding is available in the files \texttt{DEAK.thy} and \texttt{DEAK\_core.thy}. The latter contains the encoding of the terms of D.EAK, which differs only slightly from the one of the shallow embedding. It also contains functions match and replace, plus some easy lemmas about their behaviour. The functions match and replace are used  in \texttt{DEAK.thy} to define how rules are applied to sequents.

\texttt{DEAK.thy} starts out by defining the datatypes \texttt{Rule} and \texttt{Prooftree}. The function \texttt{der} implements how to reason backwards from a goal:
\begin{pyglist}[language=isabelle]
fun der :: "Locale ⇒ Rule ⇒ Sequent ⇒ (Rule * Sequent list)"
\end{pyglist}
\noindent takes a locale, a rule $r$, and a sequent $s$ and outputs the list of premises needed to prove the $s$ via $r$.%
\footnote{It is at this point where our implementation of the deep embedding is currently tailored towards substructural logics: For each rule $r$ and each sequent $s$, there is only one list of premises to consider. Generalising the deep embedding to sequent calculi with rules such as \eqref{eq:or-seq} would require a modification: If we interpret the structure $W,X,A\vee B$ in \eqref{eq:or-seq} not as a structure (ie tree) but as a list, then matching the rule \eqref{eq:or-seq} against a sequent would typically not determine the sublists matching $W$ and $X$ in a unique way. More information is available at \cite{Balco:tutorial}.} 
This function is then used to define the predicate \texttt{isProofTree} and other functions that are used by the UI.

One reason to define the deep embedding in Isabelle (and not e.g. directly in the UI) is that we want to use it in future work to implement, and prove correct, cut elimination for D.EAK and related calculi.

\subsection{Functionality of the user interface (UI)}

For the reasons described at the end of Section~\ref{sec:deak}, the UI is an essential part of the tool. The UI provides the following functionality:
\begin{itemize}
\item
    LaTeX typesetting of the terms of the calculus, with user specified syntactic sugar.
\item
    Graphical representation of proof trees in LaTeX
\item
    Exporting proof trees to LaTeX/pdf and to Isabelle (both SE and DE).
\item
    Automatic proof search (to a modest depth of 5).
\item
    Interactive proof tree creation and modification, including merging proof trees, deleting portions of proof trees, and applying rules. 
\item Tactics for deriving the generalised identity and atom rules.
\item User defined abbreviations and macros (derived rules). 
\end{itemize}
\noindent
The UI is implemented in Scala. There were several reasons for choosing Scala, one of which is Isabelle's code export functionality which translates functions written in Isabelle theory files to be exported into functional languages such as Scala or Haskell, amongst others. This meant that the underlying formalisation of terms, rules and proof trees of the deep embedding of the calculus and the functions necessary for building and verifying proof trees could be built in Isabelle and then exported for the UI into Scala.

Another advantage of using Scala is the fact that it is based on Java and runs on the JVM, which makes code execution fast enough, and, more importantly,  is cross platform and allows the use of Java libraries. This was especially useful when creating the graphical interface for manipulating proof trees, as the UI depends on two libraries, JLaTeXMath
and abego TreeLayout%
, which allow for easy typesetting and pretty-printing of the proof trees as well as simple visual creation and modification of proof trees in the UI.

\section{Case study: The muddy children puzzle}\label{sec:mc}

The muddy children puzzle is a classical example of reasoning in dynamic epistemic logic, since it highlights how epistemic actions such as public announcements modify the knowledge of agents. We will recall the puzzle  in some detail below. The solution will state that, after $k$ rounds of all agents announcing ``I don't know'', all agents do in fact know.

The correctness of the solution has been established, for all $k\in\mathbb{N}$ using induction,  by informal mathematical proof \cite{FHMV} and by mathematical proofs about a formalisation in a Hilbert calculus \cite{MaPS14}. It has also been automatically verified, for small values of $k$, using techniques from model checking \cite{HalpernVardi}  and automated theorem proving \cite{Dyckhoff,Truffaut:msc}.

Here, we prove in Isabelle/HOL that for all $k$ the solution is derivable in D.EAK.

\subsection{The muddy children puzzle}
There are $n>0$ children and $0 < k \leq n$ of them have mud on their foreheads. Each child sees (and hence knows) which of the others is dirty. But they cannot see (and therefore do not know at the beginning) whether they are dirty themselves (thus the number $n$ is known to them but $k$ is not). The first epistemic action is the father announcing (publicly and truthfully) that at least one of the children is dirty. From then on the protocol proceeds in rounds. In each round all children announce (simultaneously, publicly, truthfully) whether they know that they are dirty or not. How many rounds need to be played until the dirty children know that they are dirty?

\newcommand{\father}{\textsf{father}}
\newcommand{\no}{\textsf{no}}

\medskip\noindent In case $n=1,k=1$ the only child knows that it must be dirty, since the announcement by the father, as all announcements in this protocol, are assumed to be truthful. We write this as 
$$[\father]\Box_1 D_1,$$
 where $D_j$ is an atomic proposition encoding that child $j$ is dirty, $\Box_j p$ means child $j$ knows $p$ and $[\father]p$ means that $p$ after father's announcement.

\medskip\noindent The case $n>1$ and $k=1$ is similar. Let $j$ be the dirty child. It sees, and therefore knows, that all the other children are clean. Since, after father's announcement, child $j$ knows that there is at least one dirty child, it must be $j$, and $j$ knows it.

\medskip\noindent In case $n>1$ and $k=2$ let $J=\{j,h\}$ be the set of dirty children. After father's announcement both $j$ and $h$ see one dirty child. But they do not know whether they are dirty themselves. So, according to the protocol, they announce that they do not know whether they are dirty. From the fact that $h$ announced $\neg\Box_h D_h$, child $j$ can conclude $D_j$, that is, we have $\Box_j D_j$.
To see this, $j$ reasons that if $j$ was clean, then $h$ would be in the situation of the previous paragraph, that is, we had $\Box_h D_h$, in contradiction to the truthfulness of the announcement of $h$.
Summarising, we have shown 
$$[\father][\no]\Box_j D_j,$$
where $[\no]$ is the modal operator corresponding to the children announcing that they don't know whether they are dirty. 

\medskip\noindent The cases for $k>2$ follow similarly, so that we obtain for all dirty children $j$
\begin{equation}\label{eq:fathernoboxdj}
[\father][\no]^{k-1}\,\Box_j D_j
 \end{equation}
For example, for $n=k=100$, after 99 rounds of announcements  ``I don't know whether I am dirty'' by the children, they all do know that they are dirty. 

\subsection{Muddy children in Isabelle}
Our proof in Isabelle follows \cite[Prop.24]{MaPS14}, which gives a mathematical proof that for all $n,k>0$ there is, in a Hilbert system equivalent to D.EAK, a derivation  of 
\eqref{eq:fathernoboxdj} from the assumption
\newcommand{\dirty}{\textsf{dirty}}
\newcommand{\E}{\textsf{E}}
\newcommand{\vision}{\textsf{vision}}
\begin{gather}\label{eq:mc-assumptions}
\dirty(n,J) \ \wedge \ \E(n)^k(\vision(n))
\end{gather}
which encodes the rules of the protocol. Specifically, $\dirty(n,J)$ encodes for each $J\subseteq\{1,\ldots n\}$ that precisely the children $j\in J$ are dirty, $\vision(n)$ expresses that each child knows whether any of the other children are dirty, 
$\E(n)(\phi)$ means that `every one of the $n$ children knows $\phi$' and $f^k$ indicates $k$-fold integration of the function $f$ so that $\E(n)^k(\vision(n))$ says that `each child knowing whether the others are dirty' is common knowledge up to depth $k$.

\medskip\noindent This means that we need to prove by induction on $n$ and $k$ that for  all $n,k$ there is a derivation in the calculus D.EAK of the sequent 
\begin{equation}\label{eq:mc}
 \dirty(n,J) , \E(n)^k(\vision(n))\vdash [\father][\no]^{k-1}\,\Box_j D_j \ .
 \end{equation}
where the actions $\father$ and $\no$ also depend on the parameter $n$.

\medskip\noindent For the cases $k=1,2$ the proofs can be done with a reasonable effort in the UI of the tool, filling in all the details of the proof of \cite{MaPS14}. 	

\medskip\noindent But as a propositional calculus, D.EAK does not allow us to do  induction. Therefore we use the shallow embedding of D.EAK  and do the induction in the logic of Isabelle. The expressions $\dirty(n,J)$ and $\E(n)^k(\vision(n))$ and $[\father][\no]^{k-1}\,\Box_j D_j$ then are Isabelle functions that map the parameters $n,k$ to formulas (in the shallow embedding) of D.EAK, see the file \texttt{muddy-children.thy} \cite{Balco:MC}.

\medskip\noindent The first part of  \texttt{muddy-children.thy} contains the definitions of the formulas discussed above and establishes some of their basic properties. The actual proof is given as lemma \texttt{dirtyChildren}. We have taken care to follow \cite{MaPS14} closely, so that the proof of its Proposition~24 can be read as a high-level specification of the proof in Isabelle of lemma \texttt{dirtyChildren}. 

\medskip\noindent The proof in \texttt{muddy-children.thy} differs from its specification in \cite{MaPS14} only in a few minor ways. Instead of assuming the axiom of introspection $[\aga]p\to p$, we added the corresponding structural rules to the calculus. This seems justified as it is a fundamental property of knowledge we are using and also illustrates a use of modularity. Instead of introducing separate atomic propositions for dirty and clean, we treat clean as an abbreviation for not dirty, which relieves us from axiomatising the relationship between dirty and clean explicitly. But if we want an intuitionistic proof, we need to add to our assumptions that `not not dirty' implies dirty.

\subsection{Conclusions from the case study}
It took approximately 4 person-weeks to implement the proof of \cite[Prop.24]{MaPS14} in Isabelle. Part of this went into providing some `infrastructure' contained in the files \texttt{NatToString.thy} and \texttt{DEAKDerivedRules.thy} that could be reused for other case studies. On the other hand, we should say that it took maybe half a year to learn Isabelle and we couldn't have learned it from documentation and tutorials alone. At crucial points we profited from expert advice by Thomas Tuerk, Tom Ridge and Christian Urban.

For the construction of the proof in Isabelle, we made extensive use of the UI. Large parts of the Isabelle proof were constructed in the UI and exported to Isabelle.

One use one can make of the formal proof is to investigate which proof principles are actually needed. For example, examining the proof in \texttt{muddy-children.thy}, it is easy to establish that the only point where a non-intuitionistic principle is used is to prove $\neg\neg D_j\to D_j$. Instead we could have added this formula (which only says that ``not clean implies dirty'') to the logical description of the puzzle \eqref{eq:mc-assumptions}.

It may be worth pointing out that this analysis is based on the substructural analysis of classical logic on which D.EAK is built. In accordance with the principle of modularity discussed in the introduction, a proof in D.EAK is intuitionistic if and only 
if it does not use the so-called Grishin rules \texttt{Grishin\_L} and \texttt{Grishin\_R} (as defined in \texttt{DEAK.json}). Thus a simple text search for `Grishin' in \texttt{muddy-children.thy} suffices.

\section{Building your own calculus tool}\label{sec:meta-tool}
As discussed in Section~\ref{sec:deak-tool}, the DEAK toolbox consists of a set of Isabelle theory files that formalize the terms and encode the rules of this calculus, providing a base for reasoning about the properties of the calculus in the Isabelle theorem prover. The toolbox also includes a UI for building proof trees in the calculus.

On top of this, we provide the calculus toolbox, a meta-toolbox, which consists of a set of scripts and utilities used for maintaining and modifying the DEAK calculus tool and for building your own calculus tool.


The main component of the meta-toolbox is the build script, which takes in a description file of the terms and the rules of the calculus and expands this concise definition into multiple Isabelle theories and Scala code. Due to this centralised definition of the calculus, adding rules or logical connectives becomes much easier, as any changes made to the calculus affect multiple Isabelle and Scala files. The meta-toolbox thus allows for a more structured and uniform maintenance of the different encodings along with the UI.

A detailed documentation \cite{Balco:CT} and tutorial \cite{Balco:tutorial} is available.

\subsection{Describing a calculus}
We highlight some elements of how to describe a calculus such as D.EAK in the format that can be read by the calculus toolbox. 

The calculus is described in a file using the JavaScript Object Notation (JSON), in our example \texttt{DEAK.json}. This file specifies the types (Formula, Structure, Sequent, \dots), the operational and structural connectives, and the rules. For example, linking up with the discussion in Section~\ref{sec:deak-tool}, in

\begin{pyglist}[language = json]
"Sequent": {
  "type" : ["Structure", "Structure"],
  "isabelle" : "_ \\<turnstile> _",
  "ascii" : "_ |- _",
  "latex" : "_ {\\ {\\textcolor{magenta}\\boldsymbol{\\vdash}\\ } _",
  "precedence": [311,311,310]
\end{pyglist}

\noindent\texttt{"type"} specifies that a sequent consists of two structures.%
\footnote{The presence of the \texttt{\textbackslash\textbackslash} instead of just one \texttt{\textbackslash} is unfortunate but \texttt{\textbackslash} is a reserved character that needs to be escaped using \texttt{\textbackslash}.}
The next three lines specify how sequents will be typeset in Isabelle, ASCII and LaTeX. To make proofs readable in the UI, it is important that the user can specify bespoke sugared notation using, for example, LaTeX commands such as colours and fonts. 

Next we explain how rules are encoded. The encoding is divided into two parts. In the first part, under the heading \texttt{"calc\_structure\_rules" } the rules are declared. For example, we find 

\begin{pyglist}[language = json]
"Or_L" : {
  "ascii" : "Or_L",
  "latex" : "\\vee_L"
}
\end{pyglist}
\noindent telling us how the names of the rule are typeset in ASCII and LaTeX. The rule \eqref{or-deak} itself is described in the second part under the heading \texttt{"rules"} by
\begin{pyglist}[language = json]
"Or_L" : ["F?A \\/ F?B |- ?X ; ?Y", "F?A |- ?X", "F?B |- ?Y"],
\end{pyglist}
\noindent the first sequent of which is the conclusion, the following being the premises of the rule. The \texttt{?} has been defined in \texttt{DEAK.json} to indicate the placeholders (aka free variables or meta-variables) that are instantiated when applying the rule. The \texttt{F} marks placeholders that can be instantiated by formulas only.

The description of \texttt{Or\_L} above suffices to compile it to Isabelle. But some rules of D.EAK need to be implemented subject to restrictions expressed separately. For example the so-called atom rule formalises that in D.EAK actions do not change facts (but they may change knowledge). Thus, whereas the rule is encoded as 

\begin{pyglist}[language = json]
"Atom" : ["?X |- ?Y", ""],
\end{pyglist}
\noindent we need to enforce the condition that \texttt{?X |- ?Y} is of the form $\Gamma p \vdash \Delta p$, where $p$ is an atomic proposition and $\Gamma,\Delta$ are strings of action modalities. This is done by noting in the calculus description file the dependence on a condition called \texttt{atom} as follows.

\begin{pyglist}[language = json]
"Atom" : {
  "ascii" : "Atom",
  "latex" : "Atom",
  "condition" : "atom"			
},
\end{pyglist}
\noindent The condition itself is then implemented directly in Isabelle. 

For bottom-up proof search, the deep embedding provides a function that, given a sequent and a rule, computes the list of premises (if the rule is applicable). For the cut rule, this is implemented by looking for a cut-formula in the corresponding \texttt{Locale}, see Section~\ref{sec:se}.

\begin{pyglist}[language = json]
"RuleCut" : {
 "SingleCut" : {
  "ascii" : "Cut",
  "latex" : "Cut",
  "locale" : "CutFormula f",
  "premise" : "(\\<lambda>x. Some [((?\\<^sub>S ''X'') ...",
  "se_rule" : "l \\<turnstile>d (X \\<turnstile>\\<^sub>S f ..."
  }
 },
\end{pyglist}
\noindent After \texttt{"premise"} we find the Isabelle definition of the DE-version of the rule and after \texttt{"se\_rule"} the SE-version of the rule.

The most complicated rules of D.EAK are those  which describe the interaction of action and knowledge modalities and we are not going to describe them here. They need all of the additional components \texttt{condition}, \texttt{locale}, \texttt{premise}, 
\texttt{se\_rule}, to deal with side conditions which depend on actions being agent-labeled relations on actions.

The ability to easily change the calculus description file will be useful in the future, but also appeared already in this work. Compared to the version of D.EAK from \cite{DEAK}, we noticed during the work on the muddy children puzzle that we wanted to add rules \texttt{Refl\_ForwK} expressing $[\aga]p\to p$ (i.e.\ that the knowledge-relation is reflexive) and rules \texttt{Pre\_L} and \texttt{Pre\_R} allowing us to replace in a proof the constant representing the precondition of an action by the actual formula expressing the precondition.

\subsection{The build script, the template files, and the watcher utility}
To build the tool from the calculus description file \texttt{DEAK.json}, one runs the Python script, passing the description file to the script via the \texttt{-{}-calculus} flag. This produces the Isabelle code for the shallow and deep embedding and the Scala code for the UI. By default, this tool-code is output to a directory called \texttt{gen\_calc}. 

\medskip\noindent\textbf{Template files. } The tool-code is generated from both the calculus description file and template files. Template files contain the code that cannot be directly compiled from the calculus description file, for example, the code of the UI. But whereas the code of the UI, in the folder
\texttt{gui}, is 
independent of the particular calculus, the parser \texttt{Parser.scala} and the print class \texttt{Print.scala} consist of code written by the developer as well as code automatically generated from the calculus description file. Similarly, whereas parts of \texttt{DEAK.thy} are compiled from the calculus description file, other parts, such as the lemmas and their proofs are written by the developer.

\medskip\noindent\textbf{The Isabelle and Scala builder. } In order to support the weaving of automatically generated code into the template files, there are two domain specific languages defined in the files \texttt{isabuilder.py} and \texttt{scalabuilder.py}. For example, in the template file \texttt{Calc\_core.thy}, from which \texttt{DEAK.thy} is generated, the line

\begin{pyglist}[language = isabelle]
(*calc_structure*)
\end{pyglist}

\noindent prompts the build script to call a method defined in \texttt{isabuilder.py} which inserts the Isabelle definition of the terms of the calculus into \texttt{DEAK.thy}.

\medskip\noindent\textbf{The watcher utility. } In order to make the maintenance of the template files easier there is a watcher utility which allows, instead of directly modifying the template files, to work on the generated code. For example, if we want to change how proof search works, we would make the changes to the Isabelle file \texttt{DEAK.thy} and not directly to the template file \texttt{Calc\_core.thy}. The watcher utility, when launched, runs in the background and monitors the specified folder. Any changes made to a file inside this folder are registered and the utility decompiles this file back into its corresponding template, each time a modification occurs. The watcher utility decompiles a file by looking for any sections of the file that have been automatically generated, and replacing these definitions by the special comments that tell the build script where to put the auto-generated code. In order for the decompiling to work correctly, the auto-generated code must be enclosed by special delimiters. Looking back at the example of \texttt{(*calc\_structure*)}, when the template file is processed by the build script and expanded with the definitions from a specific calculus description file, the produced code is enclosed by the following delimiters:

\begin{pyglist}[language = isabelle]
(*calc_structure-BEGIN*)
auto-generated code ...
(*calc_structure-END*)
\end{pyglist}

\noindent
Hence, when the watcher utility decompiles a file into a template, it simply replaces anything of the form \texttt{(*<identifier>-BEGIN*) ... (*<identifier>-END*)} by the string \texttt{(*<identifier>*)}.

\section{Conclusions}\label{sec:conclusion}
We find that the tool already makes a valuable contribution to our own, so far largely theoretical research. The main directions of future work consist in extending the tool to support more ambitious projects on the proof theory as well as on the application side. These may include the following concrete projects.

\begin{itemize}
\item Proving more theorems in and about D.EAK:
\begin{itemize}
\item  case studies similar to muddy children but with dynamic updates more complicated than public announcement,
\item proving cut elimination of D.EAK in Isabelle.
\end{itemize}
\item Treating the multi-type version of D.EAK \cite{Multitype}.
\item Extending to other calculi. In particular, our methodology naturally applies to fragments of classical logics such as intuitionistic and linear logics which have many applications in computer science. 
\item Providing an interface for the calculus description file, possibly supporting the integration of different calculi.
\item Making the tool more powerful by
\begin{itemize}
\item extending available tactics,
\item improving the currently very rudimentary automated proof search.
\end{itemize}
\item Prove bigger case studies of more substantial epistemic protocols. How will the tool need to change to scale it up to bigger applications?
\end{itemize}

\noindent
Because in this paper we were interested in studying calculi such as D.EAK 
the propositional part about dynamic and epistemic operators and the higher order part of Isabelle are strictly separated, interfaced by the tool-generated shallow embedding. One should investigate building the dynamic and epistemic part directly into the higher-order logic of Isabelle. This would allow us to formalize properties such as  ``after $m$ rounds of $\no$ each child knows that there are at least $m$ dirty children'' directly instead of encoding them as functions that map parameters such as $m$ to formulas of propositional logic. On the other hand, one will loose information coming from the detailed understanding of a propositional, substructural, modular
display calculus such as D.EAK.

\medskip Possibly the most important topic of further research concerns the fact that display calculi are not directly suitable for automatic proof search. On the other hand they have the advantages of modularity we discussed. So the question is whether we can---along the lines of \cite{GorePT09}---go from display calculi constructed according to a clear proof theoretic methodology to deep inference calculi well suited for proof search. That automatic proof search is a direction worth pursuing for dynamic epistemic logics has been shown in \cite{Truffaut:msc}: for the calculus of \cite{Dyckhoff} a depth first search augmented with simple heuristics was able to automatically find a proof of the muddy children puzzle for up to 4 dirty children, see \cite[\textsection 6.4.3]{Truffaut:msc}.

\medskip Another important problem concerns how to integrate common knowledge. 
This is a well-known difficult problem. Some proposed solutions use infinitary rules, other use finitary rules are non-standard and non-modular. We plan to extend D.EAK with common knowledge while keeping it modular.

\newpage
\bibliographystyle{abbrv}

\begin{thebibliography}{10}

\bibitem{mlhandbook:BL}
F.~Baader and C.~Lutz.
\newblock {Description Logic}.
\newblock In {\em Handbook of Modal Logic}. 2006.

\bibitem{Balco:tutorial}
S.~Balco.
\newblock Building a sequent calculus toolbox.
\newblock Available at
  \url{http://goodlyrottenapple.me/2015/09/02/sequent-tutorial/}.

\bibitem{Balco:CT}
S.~Balco.
\newblock The calculus toolbox.
\newblock Download and documentation at
  \url{https://github.com/goodlyrottenapple/calculus-toolbox}.

\bibitem{Balco:DEAK}
S.~Balco.
\newblock The {DEAK} calculus tool.
\newblock Download and documentation at
  \url{https://github.com/goodlyrottenapple/DEAK-calculus-tool}.

\bibitem{Balco:MC}
S.~Balco and S.~Frittella.
\newblock Muddy children.thy.
\newblock Isabelle 2015 theory file available at
  \url{https://github.com/goodlyrottenapple/muddy-children}.

\bibitem{bms}
A.~Baltag, L.~S. Moss, and S.~Solecki.
\newblock The logic of public announcements, common knowledge and private
  suspicious.
\newblock Technical Report SEN-R9922, CWI, Amsterdam, 1999.

\bibitem{Belnap}
N.~Belnap.
\newblock Display logic.
\newblock {\em Journal of Philosophical Logic}, 11:375--417, 1982.

\bibitem{mlhandbook:BB}
P.~Blackburn and J.~van Benthem.
\newblock Modal logic: A semantic perspective.
\newblock In {\em Handbook of Modal Logic}. 2006.

\bibitem{mlhandbook}
P.~Blackburn, J.~van Benthem, and F.~Wolter, editors.
\newblock {\em Handbook of Modal Logic}.
\newblock Elsevier, 2006.

\bibitem{DawsonG01}
J.~E. Dawson and R.~Gor{\'{e}}.
\newblock Embedding display calculi into logical frameworks: Comparing twelf
  and isabelle.
\newblock {\em Electr. Notes Theor. Comput. Sci.}, 42:89--103, 2001.

\bibitem{DawsonG02}
J.~E. Dawson and R.~Gor{\'{e}}.
\newblock Formalised cut admissibility for display logic.
\newblock In {\em Theorem Proving in Higher Order Logics, 15th International
  Conference, TPHOLs 2002, Hampton, VA, USA, August 20-23, 2002, Proceedings},
  pages 131--147, 2002.

\bibitem{Dyckhoff}
R.~Dyckhoff, M.~Sadrzadeh, and J.~Truffaut.
\newblock Algebra, proof theory and applications for an intuitionistic logic of
  propositions, actions and adjoint modal operators.
\newblock {\em ACM Transactions on Computational Logic}, 14(4), 2013.

\bibitem{FHMV}
R.~Fagin, J.~Y. Halpern, Y.~Moses, and M.~Y. Vardi.
\newblock {\em Reasoning About Knowledge}.
\newblock MIT Press, 1995.

\bibitem{mlhandbook:F}
M.~Fitting.
\newblock {Modal Proof Theory}.
\newblock In {\em Handbook of Modal Logic}. 2006.

\bibitem{PDL}
S.~Frittella, G.~Greco, A.~Kurz, and A.~Palmigiano.
\newblock Multi-type display calculus for propositional dynamic logic.
\newblock DOI:10.1093/logcom/exu064, 2014.

\bibitem{Multitype}
S.~Frittella, G.~Greco, A.~Kurz, A.~Palmigiano, and V.~Sikimi\'{c}.
\newblock A multi-type display calculus for dynamic epistemic logic.
\newblock DOI:10.1093/logcom/exu068, 2014.

\bibitem{DEAK}
S.~Frittella, G.~Greco, A.~Kurz, A.~Palmigiano, and V.~Sikimi\'{c}.
\newblock A proof-theoretic semantic analysis of dynamic epistemic logic.
\newblock {\em Journal of Logic and Computation}, 2015.
\newblock DOI:10.1093/logcom/exu063.

\bibitem{Gore96}
R.~Gor{\'e}.
\newblock On the completeness of classical modal display logic.
\newblock {\em In H Wansing, editor, Proof Theory of Modal Logic}, volume 2 of
  Applied Logic:137--140, 1996.

\bibitem{Gore98}
R.~Gor\'e.
\newblock Substructural logics on display.
\newblock {\em Logic Journal of IGPL}, 6(3):451--504, 1998.

\bibitem{GorePT09}
R.~Gor{\'{e}}, L.~Postniece, and A.~Tiu.
\newblock Taming displayed tense logics using nested sequents with deep
  inference.
\newblock In {\em Automated Reasoning with Analytic Tableaux and Related
  Methods, 18th International Conference, {TABLEAUX} 2009, Oslo, Norway, July
  6-10, 2009. Proceedings}, pages 189--204, 2009.

\bibitem{HalpernVardi}
J.~Y. Halpern and M.~Y. Vardi.
\newblock Model checking vs. theorem proving: {A} manifesto.
\newblock In {\em Proceedings of the 2nd International Conference on Principles
  of Knowledge Representation and Reasoning (KR'91). Cambridge, MA, USA, April
  22-25, 1991.}, pages 325--334, 1991.

\bibitem{mlhandbook:HHSS}
I.~Horrocks, U.~Hustadt, U.~Sattler, and R.~Schmitt.
\newblock {Computational Modal Logic}.
\newblock In {\em Handbook of Modal Logic}. 2006.

\bibitem{Kracht}
M.~Kracht.
\newblock Power and weakness of the modal display calculus.
\newblock In {\em Proof theory of modal logic}, pages 93--121. Kluwer, 1996.

\bibitem{Lescanne06}
P.~Lescanne.
\newblock Mechanizing common knowledge logic using {COQ}.
\newblock {\em Ann. Math. Artif. Intell.}, 48(1-2):15--43, 2006.

\bibitem{Lescanne13}
P.~Lescanne.
\newblock Common knowledge logic in a higher order proof assistant.
\newblock In {\em Programming Logics - Essays in Memory of Harald Ganzinger},
  pages 271--284, 2013.

\bibitem{LescanneP07}
P.~Lescanne and J.~Puiss{\'{e}}gur.
\newblock Dynamic logic of common knowledge in a proof assistant.
\newblock {\em CoRR}, abs/0712.3146, 2007.

\bibitem{MaPS14}
M.~Ma, A.~Palmigiano, and M.~Sadrzadeh.
\newblock Algebraic semantics and model completeness for intuitionistic public
  announcement logic.
\newblock {\em Annals of Pure and Applied Logic}, 165(4):963--995, 2014.

\bibitem{Truffaut:msc}
J.~Truffaut.
\newblock Implementation and improvements of a cut-free sequent calculus for
  dynamic epistemic logic, 2011.
\newblock MSc thesis, University of Oxford.

\bibitem{DitmarschK15}
H.~P. van Ditmarsch and B.~Kooi.
\newblock {\em One Hundred Prisoners and a Light Bulb}.
\newblock Springer, 2015.

\bibitem{Ditmarsch}
H.~P. van Ditmarsch, W.~van~der Hoek, and B.~Kooi.
\newblock {\em Dynamic Epistemic Logic}.
\newblock Springer, 2007.

\bibitem{Wansing}
H.~Wansing.
\newblock {\em Displaying Modal Logic}.
\newblock Kluwer, 1998.

\end{thebibliography}

\appendix

\end{document}